\documentclass[12pt,a4paper]{article}
\usepackage{amsmath}
\usepackage{graphics}

\newcommand{\p}{\vspace{6pt}\noindent}
\advance\textheight by \topskip
\begin{document}
\vspace*{-2cm}
\begin{flushright}
\end{flushright}

\vspace{0.3cm}

\begin{center}
{\Large {\bf Aspects of dual models many years ago}}\\
\vspace{1cm} {\large  E.\ Corrigan\footnote{\noindent\
E-mail: {\tt edward.corrigan@durham.ac.uk} }}\\
\vspace{0.3cm} \em Department of Mathematical
Sciences\\ Durham University\\ Durham DH1 3LE, U.K.\\

\end{center}
\vskip 1cm
\begin{center}{ Abstract}\\
\bigskip
Invited contribution to the collection of articles: `The Birth of String Theory', edited by 
Andrea Cappelli, Elena Castellani, Filippo Colomo and Paolo Di Vecchia.
\end{center}

\vskip 1cm \noindent{\bf Preamble}

\p
After all this time I do not trust my memory to be accurate in every detail. Besides, I can
only provide glimpses from the perspective of a beginning graduate student struggling to
catch up followed by that of a postdoc who found himself quite far from much of the action.
I will not
even attempt to be comprehensive in either the telling or the references; rather I will restrict myself
to those aspects I concentrated on at the time, some of the people I knew and who influenced me, and to
trying not to add any insights that have emerged
with the benefit of hindsight. The latter may be hard to achieve, perhaps impossible.


\vskip .3cm \noindent{\bf Student in Cambridge}

\p
In October 1969, almost forty years ago I started my three years as a graduate student in the
Department of Applied Mathematics and Theoretical Physics, Cambridge. During the previous
year, as was customary then, I took Part III of the Mathematics Tripos.
This was a thorough grounding in many of the tools then useful in elementary particle theory.
However, by today's standards, it was  lacking in some respects, especially in the area of
quantum field theory - the most notable omission being
any mention of Yang-Mills gauge theory. This was hardly surprising, of course, because most of the
Cambridge group had been actively developing S-matrix theory and `Eden, Landshoff, Olive and
Polkinghorne' was mandatory reading for prospective graduate students. Less surprising,
was the omission of any reference to the Veneziano model, it being far too much of a recent
innovation to make it into Part III. Instead, I heard about it in a roundabout way. On the other hand,
there were compensations, such as attending Dirac's lectures on quantum mechanics, Goldstone's
on statistical mechanics and Sciama's on general relativity. Though Skyrme had already
written papers that would become influential, they were not mentioned; conformal field theory
as we now know it was far in the future.

\p
During the
Summer of 1969 I spent two months with an experimental group at the Rutherford Laboratory. The
group was led by Geoff Manning and it was my entry into the world of detectors (in fact quite large
spark chambers), and various other interesting pieces of electronics and machinery. I arrived
during a period between data taking when the group was reconfiguring its detectors, and I was
fascinated by the whole
setup. My memory tells me the purpose of the experiments was to establish various decay rates
for kaons, but I might be remembering wrongly; in any case, my interest was more in the
engineering than the physics. During the time I had between tasks in the laboratory or
experimental hall, I was learning about kaon physics, and had my first experience with FORTRAN;
but, to be frank, the intricacies of the experimental setup appeared much
more interesting than the theory. Then something unexpected happened: several members of the
group went off to a conference at Lund and I remember one member of the group, John Field,
coming back and saying to me that what I should really start to learn about was the Veneziano
model \cite{Veneziano68}.
He was enthusiastic about it because  there was a chance of real progress in understanding the
strong interactions.
As  a result, I found the paper and tried to work through  it but I was missing too much background to
really appreciate what I read.

\p
There was a lot to learn besides the prerequisites in Regge theory, duality, current algebra and so on, and,
in reality, I came to these topics after learning about the more recent developments.
Earlier in 1969, Koba and Nielsen had developed their elegant
representation of multi-particle amplitudes \cite{Koba:1969kh} and Fubini and Veneziano developed the vertex operator
representation \cite{Fubini:1969qb}, which was so useful for analyzing the spectrum. A little later, but still in 1969,
around Christmas, I seem to remember, these were followed
by Virasoro's realization that
potentially unphysical states were removed by a set of gauge conditions satisfying the beautiful
and seemingly simple algebra that now bears his name \cite{Virasoro:1969zu}.
Each of these ideas was entirely new to me and the process of placing them in order and perspective
was exceedingly daunting (as doubtless it was for almost everyone else at the time). John Polkinghorne,
the then
leader of the Cambridge elementary particle theory group in DAMTP helped  by giving a set of
introductory lectures during my first year (followed later by David Olive who gave a more technical set
of lectures on the operator formalism backed up  by the review \cite{Alessandrini:1971fx},
which was circulating in preprint form). My research supervisor
to begin with was Ian Drummond replaced after a while by David Olive who had been on leave visiting CERN
when I started (and he subsequently left Cambridge for CERN in July 1971). I do not now remember all the students
in those years. However, I did come to know
several people some of whom became friends, colleagues or co-workers later on. In the year ahead of me there was John Cardy,
David Collop and Jon Ellis and two years ahead, David Campbell, Peter Goddard, Alan White and
Wojtek Zakrzewski. Given the labyrinthine design of DAMTP we only rarely met - at coffee time or when
preprints needed assembling.

\p
During 1969 a series of preprints from Stanley Mandelstam appeared \cite{Mandelstam:1970dr}
and these were suggested to me as a suitable focus for my first investigations. The thrust of the
papers was an attempt to marry the quark model of elementary particles to the dual model, and the main
difficulty, as I soon realized - given the dual model seemed to be geared to a description of mesons -
was how to build S-matrix amplitudes involving three-quark states, or baryons. This was definitely an
important question to
tackle but it was not at all clear then (or perhaps even now) how best to do it. One of the sticking points
was the inability at that time to describe dual amplitudes for particles with spin 1/2. I also wondered about
that for a while but the solution to the problem turned out to need significantly new ideas,
which will be mentioned below. Despite being
attracted to Mandelstam's ideas, most of my time
was spent during the years 1969-71 trying to catch up and I did not make any identifiable progress. It
was a time when articles arrived at a huge rate and it was hard to keep abreast.
In fact, it was quite a depressingly slow process and I became very discouraged, coming very close to
quitting altogether. The general thinking appeared to be that dual
theory was going to be something new that had to be constructed from the `bottom up', and there was
tremendous activity, trying on the one hand to adapt the ideas to phenomenologically more realistic models and on the other
to render the model compatible with all the required properties of S-matrix theory, especially
unitarity. There was a strong feeling that people might be groping their way towards something
beyond quantum field theory. And string theory itself was still a couple of years in the future.
My exact contemporaries (Peter Collins,
Tony Mason and Charles Pantin) seemed
to be making faster progress on other topics and I began
to doubt the sense of trying to work in such a relatively new and fast-moving area. Eventually,
I was rescued by David Olive
who pointed me towards alternative thinking \cite{Ellis:1971sg}, and  using the Koba-Nielsen formulation.
This
approach led to results developed with Claus Montonen (who
arrived as a research student in October 1970), and
David; they were never published but became the  second chapter of my thesis.  For a while I tried to
find satisfactory vertex operator expressions for these general `baryon-meson' amplitudes but failed to do so.

\p
In early 1971, Pierre Ramond's article \cite{Ramond:1971gb} on dual fermions arrived followed shortly after by
an article by Andr\'e Neveu and John Schwarz \cite{Neveu:1971rx}.
Both of these had an impact on me, the first
more immediately than the second because I have a memory of being asked to talk about it in the
group's Journal Club, meaning
I had to read it very carefully. It seemed to me to be very interesting because the arguments put forward
for the inclusion of fermions by cleverly extending the Dirac $\gamma$-matrices to a field $\Gamma^\mu(z)$
appeared radical and compelling. The fact this was my first talk - and I was terrified of speaking in public
- detracted from my enjoyment of it, but not by much. The second paper was motivated by the need to adjust
the Veneziano model to include the pion and other types of meson that appeared at first sight to have
been excluded. The fact Neveu and Schwarz introduced a set of anti-commuting operators to achieve this
was very interesting, but seemed less compelling; after all, at that time there were many ideas for
modifying the basic amplitudes and there did not appear to be any obvious reasons why anti-commuting
operators should be part of a story involving mesons, unless the newly introduced operators were somehow
to be associated with their constituent quarks.

\p
Among the many ideas I gradually acquired was a realization of the importance within the dual model of the
group $SU(1,1)$ (pointed out originally by Gliozzi \cite{Gliozzi}), with its Lie algebra generated
by $L_{\pm 1},\ L_0$, satisfying
\begin{equation}\label{SU(1,1)}
[L_m,L_n]=(m-n)L_{m+n},\ \ m,n=-1,0,+1.
\end{equation}
This was soon followed by its mysterious generalization, the Virasoro algebra, satisfying\footnote{\ Although the
critically important additional `central term' proportional to $c$ was noticed by Joe Weis.}
\begin{equation}\label{Virasoro}
[L_m,L_n]=(m-n)L_{m+n}+\frac{c\,n(n^2-1)}{12}\,\delta_{m+n,0},\ \ m,n=0,\pm 1,\pm 2,\dots.
\end{equation}
Anyone who had studied angular momentum in quantum mechanics was aware of the finite dimensional,
unitary representations of $SU(2)$ labeled by spin. Graduate students in 1969 also knew at least
a little about the unitary representations of $SU(3)$, triplets, octets, decuplets, and so on, because of
the by then standard classification of hadrons and the quark model. However, it came as a surprise to me
to learn about the rather more intricate, infinite-dimensional, representations of a non-compact group
(or of its Lie algebra), such as $SU(1,1)$. I spent quite some time finding out about these and
rediscovering for myself some of their properties that were certainly well-known to other physicists (for example,
Barut and Fronsdal \cite{barut1965},  Clavelli
and Ramond \cite{Clavelli:1970qy}\cite{Clavelli:1971qz}, and, as I now know, to some of my student colleagues though we never
discussed it at the time), and to mathematicians.

\p
During that period one of the questions I
asked myself was the following:
which representations of \eqref{SU(1,1)} might be adapted to be representations of \eqref{Virasoro}?
An allied question concerned the construction of $N$-reggeon vertices. These were known to inherit
structure from the M\"obius group of conformal, linear fractional transformations (whose Lie algebra was also
\eqref{SU(1,1)})
\begin{equation}\label{mobius}
z\rightarrow z^\prime = \frac{az+b}{cz+d},\ ad-bc=1,
\end{equation}
and the question posed by several people, including Clavelli and Ramond, was how the representations of this
group might be exploited to construct more general amplitudes than those already known at that time.
Claus Montonen and I also worked together on this topic and it became the content of the paper
\cite{Corrigan:1972ux} (for each of us our first). The main conclusion was that the only
possible $N$-reggeon vertices that would be permitted
on the basis of satisfying all the desirable properties then required (and also ignoring spin 1/2)
were the standard one \cite{Lovelace:1970ej}
and those associated with the Neveu-Schwarz model. Moreover, we related the then mysterious use of
anti-commuting annihilation and creation operators in the latter model to a specific
representation  of $SU(1,1)$
that could be extended to the full Virasoro algebra only if anti-commuting operators were used. Actually,
it was essentially one of only two -
the other being the one given in terms of the standard commuting operators used by Virasoro.

\p
At that time, the only representations of \eqref{Virasoro} I was aware of were those associated with
quadratic expressions in annihilation and creation operators. The unitary representations of $SU(1,1)$
are labeled by two numbers $J,k$, and the ones that appeared to be relevant at first, in the sense
they could be extended to representations of the full Virasoro algebra (using just a single set of commuting
or anti-commuting annihilation and creation operators), were the representations
$(0,0)$ (Fubini-Veneziano), $(-1/2,0)$ (Ramond) and $(-1/2,1/2)$ (Neveu-Schwarz), and it is convenient to
call the commuting or anti-commuting Fock space operators $a_n^\mu$, $d_n^\mu$ and $b_{n+1/2}^\mu$,
respectively, as was  customary at the time. I also noticed there was another possibility: one could also
use the $(0,1/2)$ representation of $SU(1,1)$ and this would provide a representation of the Virasoro algebra
in terms of
commuting-type Fock space operators
$c^\mu_{n+1/2}$, also labeled by half odd integers. Apart from mentioning its existence in my thesis
I did not explore
it further because it appeared at the time to have no clear interpretation within the dual model.

\p
If the Virasoro
generators for the various
representations are referred to by $L_n^{(x)},\ x=a,b,c,d$, then the types labeled by $a,b$ and $c,d$
share some properties (though care needs to be taken with the Fock space operators
labeled by zero). For example, if the associated ground state is represented by $|0\rangle$ then
\begin{equation}\nonumber
L_n^{(x)}|0\rangle=0,\ n=0, \pm 1,\  x=a,b.
\end{equation}
But, on the other hand,
\begin{equation}\label{L-differences}\nonumber
\quad L_{1}^{(x)}|0\rangle= 0,\ \left(L_{0}^{(x)}-\frac{D}{16}\right)|0\rangle= 0,
\ L_{-1}^{(x)}|0\rangle\ne 0,\ x=c,d.
\end{equation}
Also note, to ensure the Virasoro algebra took the standard form \eqref{Virasoro} in each of the cases $c,d$
the operator $L_0$ had to be adjusted by the addition of $D/16$
(in $D$ space-time dimensions). In the Ramond model of dual fermions this was understood clearly to be
necessary
because the operators $d_0^\mu$ were to be identified with a set of Dirac $\gamma-$matrices. However,
in the extra possibility it seemed more mysterious. It was only later, and in a rather surprising way,
that the commuting half odd
integer  labeled Fock space operators found a more or less `natural' home \cite{Corrigan:1975sn}.
More will be written about that in the next section. To conclude this paragraph I also noticed, probably
some time later, the well-known identity among partition functions
\begin{equation*}
\prod_{n=1}^\infty \left(1-q^{n+1/2}\right)^{-1}=
\prod_{n=1}^\infty \left(1+q^{n+1/2}\right)\prod_{n=1}^\infty \left(1+q^{n}\right),
\end{equation*}
which suggests an equivalence between the states created by the $c^\dagger_{n+1/2}$ and those generated by the pairs
$$(b^\dagger_{n+1/2},\ d^\dagger_m).$$

\p
Returning to the Ramond and Neveu-Schwarz story for a moment, the other remarkable discovery
each made  was to extend the Virasoro algebra with a set of anti-commuting generators
($F_N$, integer-labeled for Ramond, and $G_r$, half odd integer-labeled for Neveu-Schwarz) and these were
just right, as it eventually turned out, for ensuring the removal of the additional ghost states arising from the
time components of the extra anti-commuting creation operators. Besides noting these algebras, I did not pay much
attention until the later months of 1971.

\p
In fact, as I already mentioned, David Olive returned to CERN in the Summer of 1971 and he arranged for me
to accompany him for part of that year - an exciting, though daunting, prospect - and I went for about
two months, from October to December. There, I renewed my acquaintances with John Cardy, who
had just started as a postdoc, Peter Goddard who was then in his second of two years at CERN,
and Bruno Renner, who had recently left Cambridge for a long-term appointment at CERN, and
met many other physicists for the first time. The most striking aspect of that visit was the very tangible feeling
that a substantial crowd of pioneers was really making waves. I tended to be very quiet then (and even now,
most of the time), but I attended a great many seminars and learned hugely more than I ever could in
Cambridge. It helped to put the whole subject in perspective.

\p
The visit was particularly useful because David and I,
prompted by the work of Thorn \cite{Thorn:1971jc},
started to look for a convenient expression for a `fermion emission vertex' as a preliminary to calculating
amplitudes with at least four fermions. By convenient was meant an expression that respected all the gauge
conditions and was as close as possible to being a standard vertex operator. The basic idea was that a
suitable vertex operator would have to `intertwine' the Ramond vertex for a fermion emitting a boson with
the Neveu-Schwarz vertex for a boson emitting a boson. That this was plausible was not unreasonable because both
fields were fermionic; that it would be tricky seemed likely because the two sorts of fields built out of
anti-commuting
operators seemed to belong to inequivalent representations of $SU(1,1)$ (and also of the Virasoro algebra -
recall the remarks surrounding eq\eqref{L-differences}). If $|A\rangle$ was a general fermion state (ie a spinor
constructed in the $d$  Fock space), then the fermion emission vertex $W_A(z)$ ought at least
to have the $SU(1,1)$ property
\begin{equation}\label{Wproperty1}\nonumber
O_d(\gamma)W_A(z) O_b^{-1}(\gamma)=(cz+d)^{-2c_d}\, W_A(\gamma(z)),
\end{equation}
and, setting $W_A(z)=\exp(zL^{(d)}_{-1})\,\tilde W_A(z)$, the further properties
\begin{equation}\label{Wproperty2}
\tilde W_A(z)|0_b\rangle = |A\rangle,\ \ \tilde W_A(z)\,\frac{i\sqrt{2}\gamma^{d+1}H^\mu(y)}{\sqrt{y}}=
\frac{\Gamma^\mu(y-z)}{\sqrt{y-z}}\,\tilde W_A(z),
\end{equation}
where $L^{(d)}_0|A\rangle=c_d|A\rangle$ and $m_A^2/2=c_d-1$. (Note, at the time it was not yet realised that the mass
of the lightest Ramond fermion was zero, or that the critical dimension was ten.) Great care had to be taken
with the second of the properties listed in \eqref{Wproperty2} because of the singularities on either side.
Details of the vertex
operator $\tilde W_A(z)$ and some of its other properties are to be found in \cite{Corrigan:1972tg}. While it was designed
to have the transformation property \eqref{Wproperty1} the vertex also converted (a sum of) Virasoro generators in the
$d$ Fock space to a Virasoro generator in the $b$ Fock space, a property that was sufficient to guarantee the
Virasoro gauge conditions were transmitted adequately through the vertex. At that time, the relationship between the
fermion emission vertex and the  $F_N$, $G_r$ generators was not understood. This was clarified later.

\p
The final part of my thesis involved an attempt to calculate the four-fermion scattering amplitude based
on the ingredients known up to that point. However, the missing step was learning how to remove all the
ghost states; I did not know how to do that because the fermion emission vertex did not appear to deal
correctly with the gauge operators $F_N$ and $G_r$ (that is, to convert one set into the other). I tried out
 a rudimentary projection operator onto physical states but it was  incomplete and unsatisfactory.
For that reason I did not try to publish what I had found. In fact, the calculations were published
later in \cite{Corrigan:1974hp}, which appeared, because of delays, after
the article  it was intended to precede \cite{Corrigan:1974vz}. It took some time to unfold the story of
how the two sets of gauges were interrelated, and that story belongs in the next section.

\p
To conclude this chapter: during my last months as a student I had little contact with either of my
supervisors but I remember John Polkinghorne being concerned and asking me from time to time
what I was up to. Though I was stuck on a specific issue, in the end I wrote my thesis
 intensively, and quite quickly,
and  typed up the hand-written version in a couple of weeks (my typewriter bought for this
purpose did not have any mathematical symbols so the equations remained hand-written, which was
acceptable at the time though unthinkable now).

\p
My
external examiner was Korkut Bardak\c{c}i and
fortunately he approved.

\vskip .3cm \noindent{\bf Research Fellow in Durham}

\p
During the later part of 1971 and during 1972 I was looking for a postdoc position, though I felt it was
an unlikely proposition because I had no publications at the moment I began applying.
At that time the competition to go to the United States was very fierce and there were
only limited possibilities in Europe or the UK. Actually, this was a period when the UK economy was
already in a parlous state and, though the `Three Day Week' would not be introduced until 1973, the future
looked bleak. With a young family (my first daughter had been born in January 1972), stability was needed,
and the situation made me (and I imagine many others too) very nervous. Anyway, I applied for many
jobs but, in the end,  it was what had seemed
to be the remotest of chances that brought the best prospects. David Fairlie, who at that time I knew  by
name - though
 I had attended a seminar he gave in Cambridge -
wrote to me about the A.J. Wheeler Fellowships at Durham encouraging me to apply. (These are
now called Addison Wheeler Fellowships and restricted
to areas that explicitly exclude mathematics or physics). At the time they were
open to  all; I applied, was asked for interview, and was very surprised (and delighted)
to be offered one. It was actually a very good deal because the contract could last for up to five years, yet
I had mixed feelings about returning northwards (my roots were in Birkenhead and Manchester),
when CERN seemed to be the
centre of activity. I was not really aware
of the long term goal to build up the Durham group following the appointments of David Fairlie, Alan Martin
and Euan Squires
some years previously. However, Peter Goddard, who, as I mentioned I knew a little from Cambridge and my
visit to CERN, and who I also knew had played a major role in the `no ghost' theorem and the development of
the relativistic string (with Goldstone, Rebbi and Thorn), was also moving to Durham  in October 1972. I
soon appreciated that, far from being remote, Durham was very firmly on the map. Within a few months there
had been visits by Lars Brink, Paolo Di Vecchia, Holger Nielsen and Charles Thorn, and probably a number
of others I cannot now remember who came briefly to give talks.

\p
It's always difficult moving to a new place (and I lived to begin with in Peterlee - some
distance from Durham). For a few months, but unsuccessfully, I kept on trying to think of
a way to sort out the gauge conditions for the fermion vertex and to evaluate some of the strange
functions that had appeared in what I had done so far with the four fermion scattering amplitude.
Peter was interested in the problem and we decided to look very carefully
again at the intertwining property of the fermion emission vertex to see if something had been
overlooked. We felt it necessary to adapt the formalism a bit (basically to make more precise the
intertwining property \eqref{Wproperty2}) and discovered that the best that could be done was
to arrange for a specific  Ramond anti-commuting generator $F_N, \ N>0$ to be converted to a
linear combination $\sum_s f_{Ns}G_s$ containing generators $G_s$ with both positive and negative
labels \cite{Corrigan:1973jz}. This appeared to be a negative result.

\p
Meanwhile,  Lars Brink and David Olive introduced their projection operator onto on-shell physical states
\cite{Brink:1973qm}: this brought new insights and proved to be a very useful tool.
Subsequently, Lars and David (with Claudio Rebbi and Jo\"el Scherk) noted in
\cite{Brink:1973jd} that if the fermion ground state mass was zero and the critical dimension
 was ten, then, despite the apparently strange behaviour of the gauges noted in \cite{Corrigan:1973jz},
only physical states of the
Neveu-Schwarz model coupled to a fermion-anti-fermion pair. Moreover, David and Jo\"el \cite{Olive:1974sv}
inserted the projection operator onto the Neveu-Schwarz physical states into the four fermion calculation to derive
an improved (but not evaluated) four-fermion scattering amplitude. The improvement had an interesting structure since
it involved the determinant of an infinite
dimensional matrix $\Delta(x)$ defined by
\begin{equation}\label{Delta}\nonumber
\Delta(x)={\rm det}\left(1-M^2(x)\right),
\end{equation}
with
\begin{equation}\nonumber
\ M_{rs}(x)= - (-x)^{r+s}\frac{r}{r+s}\,
\left(\begin{array}{c} -1/2 \\r-1/2 \end{array}\right)\left(\begin{array}{c}-1/2 \\r-1/2
\end{array}\right),\ r,s =1/2,3/2,\dots
\end{equation}
The evaluation of the amplitude was quite complicated and was done in stages. It seemed to be
an interesting calculation to do because it was not clear from the start that the result would have the same
spectrum of meson states in the crossed channel (and actually it turned out not to be quite the same).
The work I had done before and eventually published in \cite{Corrigan:1974hp}
was relevant yet incomplete. Nevertheless, it established several relationships between the ingredients, including
the determinant of the antisymmetric part of $M$, and formed the basis of our subsequent evaluation.
\p
Schwarz and Wu \cite{Schwarz:1973jf} first
guessed and checked numerically that $$\Delta(x)=(1-x)^{-1/4},$$ while Peter and I (with Russell
Smith\footnote{\noindent\ Russell Smith
(now retired) is a pure mathematician whose expertise in differential equations led him to suggest some tricks we
might never have thought of  otherwise.} and David Olive)
found an analytical way of evaluating all the pieces, including this and other determinants \cite{Corrigan:1974vz}.
We also
analyzed carefully the interesting manner in which the Fierz transformation in ten dimensions \cite{Case}
meshed beautifully with
the duality of the four-fermion amplitude\footnote{Intriguingly, the Fierz transformation represents the transformation
$\phi:\ z\rightarrow (1-z)/(1+z),$ with  $\phi^2=1$.}. These results reinforced Mandelstam's entirely different,
interacting string, approach to the same calculation \cite{Mandelstam:1974hk}, which appeared as a preprint
shortly before we submitted  our article.
The whole story was summarised by Jo\"el Scherk in the impressive review he wrote a few months later \cite{Scherk:1974jj}.

\p
On a side issue:
Peter and I also wrote a paper at about the same time using
the projection operator technique to
extend the `no ghost' theorem to the Ramond model, verifying  it required the zero mass condition and ten dimensions
for the physical states to be entirely transverse (as had been suggested earlier using different arguments by Thorn, Rebbi and
Schwarz) \cite{Corrigan:1974jb}.

\p
Looking back at work that was completed thirty-five years ago is quite sobering especially as one tends to see, with
the benefit of hindsight, the missed clues and mistaken paths. Nevertheless, this was a hard problem,
requiring a number of complementary steps taken by different groups of people, with  interdependent
approaches, to reach a full understanding within the technology of the time. It would be another thirteen years
before new technology enabled an alternative, and rather slicker, derivation of the same quantity \cite{Cohn:1986bn}.
One can also see how, in the days before electronic
communication, small delays might occur in the preparation and distribution of preprints, submission dates, publication dates,
and how the chronology of ideas can be affected by the pattern of conferences and personal encounters. All these factors
cause small perturbations that may be magnified with time and the provenance of some ideas are lost while others
are highlighted. Probably this is inevitable and it is perhaps better merely to look at the achievements as a whole and how they
provided some of the steps towards where we are now. I also remember, possibly in September 1973, though I can no longer
be certain precisely when, a conversation with two people in the CERN library. One of them was rather critical of the
attempts to perform these complicated calculations since it was already by that time clear that the dual model,
or even string theory, was going to have a hard time being phenomenologically acceptable; better in his view
to be doing something
else. The other, Joe Weis, was supportive, taking the view that you had best do what you can to the end because
you can never be certain in advance where it might lead.

\p
The Summer of 1974 was eventful. David, Peter and I were invited by John Schwarz to take part in the summer programme at Aspen
and this was another novel experience. Aspen was a wonderful place to spend a few weeks. Not only did I meet
many other physicists, there was always the music to accompany the seminars that were held outside.
However, I did not really have much new to add to the meeting. The
only idea I had had concerned the curious crossing properties of the four fermion vertex. It was already known
that the fermions had to be massless and I (and probably several others) noticed that the crossing properties
made more sense if the fermions were designed to be `left-handed' rather than Dirac particles.
Since we were discussing a putative model of hadrons, that seemed odd. It occurred to me these might be
regarded instead as neutrinos and I wondered for a while if one could assemble the elements of the `Standard Model'.
However, it did not really work because the Weinberg angle came out wrongly (to $ \pi/3$) and I felt that was
unsatisfactory. Apart from mentioning this in my seminar\footnote{I am grateful to Peter Goddard for showing
me recently his notes
of that seminar!} and discussing it with Peter and briefly with Jo\"el Scherk -
who I knew from my stay at CERN
and  was also at Aspen - and realising he had had the same idea, I did not pursue it further.

\p
Besides the London conference in the July, and the few intensive weeks at Aspen in August, I started
to work on an idea of David Fairlie's that he had had some time before but had never developed. His idea was to
adjust the `analogue model' \cite{Fairlie Nielsen} so that, in addition to having point sources on the rim of the disk where on-shell momentum
was injected - corresponding to on shell particles - there should also be portions of the perimeter of the disk
where momentum would be `smeared out'. These would not correspond to particles on-shell. Thus, particles would be associated
with specific locations $z_i$ on the rim of the unit disk (as in the Koba-Nielsen formulation), while off-shell momentum insertions would correspond to pairs of points $z^\prime_j, \
z_j^{\prime\prime}$.

\p
It's perhaps easiest to express the idea by thinking about it in a conformally
equivalent way with the disk
mapped to the upper half-plane and its rim lying along the real axis. From that perspective, and denoting
the complex potential by
$f^\mu(z)$,
 potentials corresponding to a set of sources along the real line should have the property that their
imaginary parts jump by $k_i^\mu$  as $z$ moves past $z_i$ along the real line when there is a point source
 at $z_i$ or, by an  amount $Q_j^\mu$ as $z$ moves from $z^\prime_j$ to $z^{\prime\prime}_j$ when the momentum
is smeared out. Moreover, in the
latter case the real parts of $f^\mu(z)$ remain constant as  $z$ moves from $z^\prime_j$ to $z^{\prime\prime}_j$.
A potential with  a single pair $z^\prime$ and $z^{\prime\prime}$ and these properties has the form
\begin{equation}\label{offshellpotential}f^\mu(z)=\sum_{i=1}^Nk_i^\mu\ln\left(
\frac{\sqrt{\frac{z_i-z^\prime}{z_i-z^{\prime\prime}}}-
\sqrt{\frac{z-z^\prime}{z-z^{\prime\prime}}}}{\sqrt{\frac{z_i-z^\prime}{z_i-z^{\prime\prime}}}+
\sqrt{\frac{z-z^\prime}{z-z^{\prime\prime}}}} \, \right).
\end{equation}
This should be compared with
$$f^\mu(z)=\sum_{i=1}^Nk_i^\mu\ln\left(z-z_i\right),$$
which leads via the analogue procedure \cite{Fairlie Nielsen} to the standard Koba-Nielsen expression for a multi-particle on-shell amplitude.
In \eqref{offshellpotential}
there is no requirement that $Q^\mu=-\sum_{i=1}^N k_i^\mu$ be on-shell. Any attempt to write a corresponding potential
with two or more strips leads to elliptic functions that are difficult, if not impossible, to write down explicitly.
We also noted that in the Mandelstam picture of interacting strings our strips corresponded to strings terminating
at finite times. Because it looked difficult to write down multiparticle amplitudes with some particles off-shell
we decided to adopt a different strategy: to factorize the amplitude constructed using \eqref{offshellpotential}
and use the factorization to identify a vertex operator describing a process in which a string emits off-shell momentum.
This was a long shot because the vertex operator had to preserve all the desirable features of the Fubini-Veneziano
vertex operator (for example its transformation properties and propagating correctly the Virasoro gauge conditions), and this
had been tried before without success (see, for example, Clavelli and Ramond \cite{Clavelli:1971qz}
). Nevertheless, we were encouraged by the close resemblance our one off-shell/many on-shell
amplitude bore to the amplitudes
written down previously by Schwarz \cite{Schwarz:1974kd}. He had started with a seemingly very different set of
assumptions - that apparently only worked properly in sixteen dimensions. It was the latter fact,
when we made the connection with
Schwarz's work, that triggered something in my mind and I suddenly remembered the (n+1/2)-labeled
bosonic annihilation and creation operators. To my surprise and delight vertex operators constructed from these
were very close to what we wanted, though by no means the whole story. Besides producing the amplitudes for us
there was the advantage that there was no place for `zero modes' because there was no momentum operator
naturally linked to the vertex.

\p
In the end we developed a reasonably coherent picture and the technicalities of the story can be found in
\cite{Corrigan:1975sn}. Here, I will merely summarize it.

\p
Our off-shell vertex
$J(z^\prime,z^{\prime\prime},Q)$ was a formal product of three pieces: first, an operator that switched (or intertwined)
the usual modes of the string (but excluding the position operator $q$) with an auxiliary set labeled by (n+1/2)-labeled
operators, second, the operator
$e^{iq\cdot Q}$ that injected an arbitrary amount of momentum into the system, and third, an operator that converted back
to the usual modes. All of this was done in a manner preserving the essential features of a vertex operator yet
needing no requirement on $Q^2$.
In stringy language one may say it more picturesquely. In order to shed or gain momentum the string first switches
to a state in which the momentum and the rest of the even modes are lost and effectively replaced
by the $(n+1/2)$-labelled modes.
 One could say, as we did at the time, that
instead of Neumann conditions at both ends of an open string one should impose a Dirichlet condition at one end
and a Neumann condition at the other,
implying that the string no longer conserved momentum - and hence momentum could be lost or gained via the
pinned down end. The mixed conditions also required the unfamiliar modes.
In other words, to go off-shell in our picture, an open string switches its boundary conditions to
Neumann-Dirchlet and back again after the momentum has been inserted or extracted. The strange condition,
that the dimension of
space time ought to be sixteen for all this to work, had an explanation of sorts in the inevitable presence of the
operators $c^\mu_r$. Besides, as we also pointed out, in the Neveu-Schwarz model this peculiarity appeared to be
ameliorated with the critical dimension restored to ten. It did not occur to us to apply the trickery to just
sixteen of the twenty-six components of the bosonic string. With hindsight, we did not explain  the idea
in the best way possible - and we ought to have pursued it further (but we were distracted by generalizations of
the 't Hooft-Polyakov monopole \cite{COFN}). At the time
it was a persistent problem that had been looked at by many people over the years from 1969; examples,
to illustrate the range of ideas but with no pretence at completeness, are given in \cite{Collins:1974bq}.

\p
To modern eyes, doubtless, the whole question looks redundant because string/brane
theory is a `model for everything' and the questions we were trying to answer (how would a hadronic system
interact with
an electromagnetic or weak current whose origins were external to itself?) have now evaporated. Moreover,
the language looks somewhat old-fashioned when compared with the streamlined arguments that are now commonplace.
Nevertheless, for quite some time afterwards I was
unreasonably elated by the picture we had uncovered, which seemed to me to be quite natural, emerging as it had
in such a round about way without being imposed at the start. Later on, Warren Siegel \cite{Siegel:1976qn} and Mike
Green \cite{Green:1977yt} took the idea up
for a while but the idea of mixing boundary conditions had to wait more than a decade before Polchinski
discovered its significance in a very different context. Some of the technology proved to be useful later in some
work \cite{Corrigan:1987tv} I did with Tim Hollowood, my then graduate student.

\p
As I mentioned at the start of this part of the story, it
began in the late Summer of 1974 but I should also say it continued as I moved to CERN to take up the CERN Fellowship
that I had been awarded
earlier in the year. It was completed there, but I remember vividly the hectic few days of transition when I
was busy calculating during the lulls in the  family move to Geneva. For what it's worth, I
have noticed several times how a physical displacement, or change of job, can facilitate  ideas.

\p
There is a minor spin-off from this story: the (0,1/2) representation of $SU(1,1)$, which seemed to be
required to enable a string to go off-shell, also provided a simple derivation of the determinant needed for
the four-fermion calculation. Personally, I found this calculation and others related to it (the details were reported in
\cite{Bruce:1975ts}) more elegant and satisfying than the earlier derivations.  Unfortunately though, the
ideas arrived too late to be an essential part of the stream.

\p
Earlier, I mentioned that the fermion emission vertex resembled an intertwiner between two inequivalent representations
of $SU(1,1)$, or of the Virasoro algebra,
and that this appeared to me to be strange from a representation theory perspective. Our off-shell vertex construction
was also strange in the same sense
because it contained a pair of intertwiners performing a similarly
impossible trick. I puzzled about this for a while but was unable to resolve the apparent paradox and set it to one side.
Fortunately,  a proper analysis resolving these issues was provided later by Peter Goddard and Roger Horsley
\cite{Goddard:1976dm}. Looked at  appropriately, each type of vertex could be understood in
group theoretic terms.

\vskip .3cm \noindent {\bf Epilogue}

\p
During the Summer of 1975 David Fairlie organised a Dual Model workshop in Durham to which around 25 people came
\footnote{This is David Fairlie's recollection, mine was originally less precise.},
including Pierre Ramond, who I had not met previously,  and several
others, especially from Japan, who I knew only via their articles. My memory tells me it was a successful week
and perhaps the last meeting of its kind. I no longer have the list of participants, or the programme,
so it is hard to be precise, and I
was distracted for part of the time. The reason for this was that
during the meeting I
was also interviewed for a lectureship at Durham (actually I think I am correct in remembering
it was the post vacated by Peter Goddard in 1974 when he moved
to Cambridge - it had been frozen for a while to cut costs), offered the job and, instantly accepted it to start in
January 1976. This meant I would have to cut short the CERN Fellowship, which was a pity, but, on the other hand, I was
very lucky: this was to be the only appointment in this area in Durham for a long time because the
UK academic system was more or less on hold owing to severe financial problems.

\p
Since then, I have, intermittently, worked with David Fairlie, Peter Goddard and David Olive on a series of problems in gauge theory,
and later, with a sequence of my own graduate students and postdocs, on various topics in integrable quantum field theory.
My time since 1976 has been divided by teaching, and increasingly by administration: I find myself
looking back with some nostalgia, and wistfully, for those few years when my main preoccupation was thinking about dual models.
 Never afterwards
has it been possible to free up such a continuous stretch of time, though I was lucky to be able to spend
a year at the Ecole Normale Sup\'erieure in Paris 1977-78, and a year at CalTech 1978-79.
There have been other periods of intensive activity, some
of which I have tried to keep up with, and there have been some personal high spots, but none of it since has
been associated with  similar feelings of excitement and common purpose. It could be this feeling is based
on an illusion arising from
my relative inexperience. That may be, but  I can also say  this: few of us are fortunate enough not to have need of illusions.

\vskip .3cm \noindent{\bf Acknowledgements} 

\p
I am indebted to Jane for her sustained support during these years, and to the other students,
especially Claus Montonen
and Charles Pantin, and grateful for the guidance and help of many other
people who kept me
going during low
moments as a graduate student - particularly Ian Drummond, David Olive and John Polkinghorne.  I am especially
grateful, to my Durham colleagues for  their kindness, enthusiasm, patient help, encouragement and collaborations, especially
David Fairlie, Peter  Goddard, Bob Johnson and Russell Smith,  and  to others I met only rarely, such as Lars Brink,
 Paolo  Di Vecchia,
Jo\"el Scherk, Pierre Ramond and Charles Thorn. I am grateful to David Fairlie, Peter Goddard and David Olive  for sharing their memories.


\vskip -1cm

\begin{thebibliography}{99}
\bibitem{Veneziano68}
G.~Veneziano,
{\it Construction of a crossing-symmetric, Regge-behaved amplitude for linearly rising trajectories},
Nuovo Cim. {\bf 57A} (1968), 190.

\bibitem{Koba:1969kh}
  Z.~Koba and H.~B.~Nielsen,
  {\it Manifestly crossing invariant parametrization of $N$ meson amplitudes},
  Nucl.\ Phys.\   {\bf B12} (1969) 517.


\bibitem{Fubini:1969qb}
  S.~Fubini and G.~Veneziano,
  {\it Level structure of dual-resonance models},
  Nuovo Cim.\   {\bf 64A} (1969) 811.

  S.~Fubini, D.~Gordon and G.~Veneziano,
  {\it A general treatment of factorization in dual resonance models},
  Phys.\ Lett.\   {\bf B29} (1969) 679.




\bibitem{Virasoro:1969zu}
  M.~Virasoro,
  {\it Subsidiary conditions and ghosts in dual resonance models},
  Phys.\ Rev.\   {\bf D1} (1970) 2933.

\bibitem{Alessandrini:1971fx}
  V.~Alessandrini, D.~Amati, M.~Le Bellac and D.~Olive,
  {\it The operator approach to dual multiparticle theory},
  Phys.\ Rep.\  {\bf 1C} (1971) 269.



\bibitem{Mandelstam:1970dr}
  S.~Mandelstam,
  {\it Relativistic quark model based on the Veneziano representation. I. meson
  trajectories},
  Phys.\ Rev.\  {\bf 184}, 1625 (1969).

  S.~Mandelstam,
  {\it 
  Relativistic  ... II. general
  trajectories},
  Phys.\ Rev.\   {\bf D1}, 1734 (1970).

  S.~Mandelstam,
  {\it 
   Relativistic  ... III. baryon
  trajectories}
  Phys.\ Rev.\   {\bf D1}, 1745 (1970).

\bibitem{Ellis:1971sg}
  S.~Ellis, P.~H.~Frampton, P.~G.~O.~Freund and D.~Gordon,
  {\it Hadrodynamics and quark structure},
  Nucl.\ Phys.\  B {\bf 24} (1970) 465.


\bibitem{Ramond:1971gb}
  P.~Ramond,
  {\it Dual theory for free fermions},
  Phys.\ Rev.\   {\bf D3} (1971) 2415.

\bibitem{Neveu:1971rx}
  A.~Neveu and J.~H.~Schwarz,
  {\it Factorizable dual model of pions},
  Nucl.\ Phys.\   {\bf B31} (1971) 86.


\bibitem{Gliozzi} F.~Gliozzi, {\it Ward-like identities and twisting operator in dual resonance models},
 Lett. Nuovo Cim. {\bf 2} (1969) 895.


 \bibitem{barut1965}
 A.~O.~Barut and C.~Fronsdal,
 {\it On non-compact groups. II. Representations of the 2+1 Lorentz Group},
Proc. Roy. Soc. {\bf A287} (1965) 532



\bibitem{Clavelli:1970qy}
  L.~Clavelli and P.~Ramond,
  {\it Group theoretical construction of dual amplitudes},
  Phys.\ Rev.\   {\bf D3} (1971) 988.

  L.~Clavelli and P.~Ramond,
  {\it SU(1,1) analysis of dual resonance models},
  Phys.\ Rev.\   {\bf D2} (1970) 973.

\bibitem{Clavelli:1971qz}
  L.~Clavelli and P.~Ramond,
  {\it New class of dual vertices}
  Phys.\ Rev.\   {\bf D4} (1971) 3098.

\bibitem{Corrigan:1972ux}
  E.~Corrigan and C.~Montonen,
  {\it General dual operatorial vertices},
  Nucl.\ Phys.\   {\bf B36} (1972) 58.


\bibitem{Lovelace:1970ej}
  C.~Lovelace,
  {\it Simple N-reggeon vertex},
  Phys.\ Lett.\   {\bf B32} (1970) 490.


  D.~I.~Olive,
  {\it Operator vertices and propagators in dual theories},
  Nuovo Cim.\   {\bf 3A} (1971) 399.






\bibitem{Thorn:1971jc}
  C.~B.~Thorn,
  {\it Embryonic dual model for pions and fermions},
  Phys.\ Rev.\  {\bf D4} (1971) 1112.



\bibitem{Corrigan:1972tg}
  E.~Corrigan and D.~I.~Olive,
  {\it Fermion-meson vertices in dual theories},
  Nuovo Cim.\  {\bf 11A} (1972) 749.

\bibitem{Corrigan:1973jz}
  E.~Corrigan and P.~Goddard,
  {\it Gauge conditions in the dual fermion model},
  Nuovo Cim.\   {\bf 18A} (1973) 339.

\bibitem{Brink:1973qm}
  L.~Brink and D.~I.~Olive,
  {\it The physical state projection operator in dual resonance models for the
  critical dimension of space-time},
  Nucl.\ Phys.\  {\bf B56} (1973) 253.

  L.~Brink and D.~I.~Olive,
  {\it Recalculation of the unitary single planar dual loop in the critical
  dimension of space time},
  Nucl.\ Phys.\  {\bf B58} (1973) 237.

  E.~Corrigan and P.~Goddard,
  {\it The off-mass shell physical state projection operator for the dual
  resonance model},
  Phys.\ Lett.\  {\bf B44} (1973) 502.


\bibitem{Brink:1973jd}
  L.~Brink, D.~I.~Olive, C.~Rebbi and J.~Scherk,
  {\it The missing gauge conditions for the dual fermion emission vertex and their
  consequences},
  Phys.\ Lett.\  {\bf B45} (1973) 379.







\bibitem{Olive:1974sv}
  D.~I.~Olive and J.~Scherk,
  {\it Towards satisfactory scattering amplitudes for dual fermions},
  Nucl.\ Phys.\ {\bf B64} (1973) 334.



\bibitem{Schwarz:1973jf}
  J.~H.~Schwarz and C.~C.~Wu,
  {\it Evaluation of dual fermion amplitudes},
  Phys.\ Lett.\   {\bf B47} (1973) 453.

  J.~H.~Schwarz and C.~C.~Wu,
 {\it Functions Occurring In Dual Fermion Amplitudes},
  Nucl.\ Phys.\  {\bf B73} (1974) 77.



\bibitem{Corrigan:1974hp}
  E.~Corrigan,
  {\it The scattering amplitude for four dual fermions},
  Nucl.\ Phys.\   {\bf B69} (1974) 325.

\bibitem{Corrigan:1974vz}
  E.~Corrigan, P.~Goddard, R.~A.~Smith and D.~I.~Olive,
  {\it Evaluation of the scattering amplitude for four dual fermions},
  Nucl.\ Phys.\   {\bf B67} (1973) 477.

\bibitem{Case}
K. M. Case, {\it Biquadratic spinor identities},
Phys. Rev. {\bf 97}(1955) 810.

\bibitem{Mandelstam:1974hk}
  S.~Mandelstam,
  {\it Interacting String Picture of the Neveu-Schwarz-Ramond Model},
  Nucl.\ Phys.\   {\bf B69} (1974) 77.



\bibitem{Scherk:1974jj}
  J.~Scherk,
  {\it An Introduction To The Theory Of Dual Models And Strings},
  Rev.\ Mod.\ Phys.\  {\bf 47} (1975) 123.

\bibitem{Corrigan:1974jb}
  E.~Corrigan and P.~Goddard,
  {\it The absence of ghosts in the dual fermion model},
  Nucl.\ Phys.\   {\bf B68} (1974) 189.


\bibitem{Cohn:1986bn}
  J.~Cohn, D.~Friedan, Z-a.~Qiu and S.~H.~Shenker,
  {\it Covariant quantization of supersymmetric string theories: the spinor field
  of the ramond-neveu-schwarz model},
  Nucl.\ Phys.\   {\bf B278} (1986) 577.

\bibitem{Fairlie Nielsen}
  D.~B.~Fairlie and H.~B.~Nielsen,
 { \it An analog model for KSV theory},
  Nucl.\ Phys.\   {\bf B20} (1970) 637.




\bibitem{Schwarz:1974kd}
  J.~H.~Schwarz,
  {\it Off-mass-shell dual amplitudes without ghosts},
  Nucl.\ Phys.\  {\bf B65} (1973) 131.

  J.~H.~Schwarz and C.~C.~Wu,
  {\it Off Mass Shell Dual Amplitudes. 2},
  Nucl.\ Phys.\   {\bf B72} (1974) 397.


\bibitem{Corrigan:1975sn}
  E.~Corrigan and D.~B.~Fairlie,
  {\it Off-shell states in dual resonance theory},
  Nucl.\ Phys.\   {\bf B91} (1975) 527.
\bibitem{COFN}
  E.~Corrigan, D.~I.~Olive, D.~B.~Fairlie and J.~Nuyts,
  {\it Magnetic monopoles in SU(3) gauge theories},
  Nucl.\ Phys.\   {\bf B106} (1976) 475.

  \bibitem{Collins:1974bq}
  Y.~Nambu,
 {\it Electromagnetic currents in dual hadrodynamics},
  Phys.\ Rev.\   {\bf D4} (1971) 1193.

  I.~T.~Drummond,
 {\it Dual amplitudes for currents},
  Nucl.\ Phys.\   {\bf B35} (1971) 269.

  H.~Sato,
 {\it Current amplitudes in the dual resonance model},
  Prog.\ Theor.\ Phys.\  {\bf 45} (1971) 1592.

  A.~Neveu and J.~Scherk,
 {\it Currents and Green's functions for dual models - I},
  Nucl.\ Phys.\   {\bf B41} (1972) 365.

P.~V.~Collins and K.~A.~Friedman,
  {\it Off-shell amplitudes and currents in the dual resonance model},
  Nuovo Cim.\  {\bf 28A} (1975) 173.

\bibitem{Siegel:1976qn}
  W.~Siegel,
  {\it Strings with dimension-dependent intercept},
  Nucl.\ Phys.\   {\bf B109} (1976) 244.

\bibitem{Green:1977yt}
  M.~B.~Green,
  {\it Dynamical point-like structure and dual strings},
  Phys.\ Lett.\  {\bf B69} (1977) 89.


  M.~B.~Green,
{\it Point-like structure and off-shell dual strings},
  Nucl.\ Phys.\   {\bf B124} (1977) 461.

\bibitem{Corrigan:1987tv}
  E.~Corrigan and T.~J.~Hollowood,
 {\it A bosonic representation of the twisted string emission vertex},
  Nucl.\ Phys.\  {\bf B303} (1988) 135.

  E.~Corrigan and T.~J.~Hollowood,
  {\it Comments on the algebra of straight, twisted and intertwining vertex
  operators}
  Nucl.\ Phys.\  {\bf B304} (1988) 77.

\bibitem{Bruce:1975ts}
  D.~Bruce, E.~Corrigan and D.~I.~Olive,
  {\it Group theoretical calculation of traces and determinants occurring in dual
  theories},
  Nucl.\ Phys.\  {\bf B95} (1975) 427.

\bibitem{Goddard:1976dm}
  P.~Goddard and R.~Horsley,
  {\it The group theoretic structure of dual vertices},
  Nucl.\ Phys.\   {\bf B111} (1976) 272.

  R.~Horsley,
  {\it The group theoretic structure of the dual current vertex},
  Nucl.\ Phys.\   {\bf B138} (1978) 474.



\end{thebibliography}
\end{document}